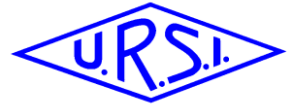

# Streaming Instability Generation in Lunar Plasma Environment

Mehul Chakraborty[1], Vipin K. Yadav[2*] and Rajneesh Kumar[1]

1 Department of Physics, Banaras Hindu University (BHU), Varanasi 221005, India

2 Space Physics Laboratory (SPL), Vikram Sarabhai Space Centre (VSSC), Thiruvananthapuram 695022, India

* Corresponding Author: vipin_ky@vssc.gov.in

## Abstract

Plasma instabilities are the non-linear processes occurring in plasmas when excess energy gets accumulated in a plasma system which is unable to hold it. There are almost 60 known plasma instabilities in nature.

## 1. Introduction

The phenomena of plasma instability takes place in a plasma where excess energy is accumulated due to the free energy sources. The plasma instabilities are often observed in plasma systems residing in space such as the Sun, the planetary ionospheres, etc. Earth's natural satellite Moon has a very thin atmosphere and subsequently feeble plasma environment. However, this tenuous plasma environment is a place of several non-linear plasma phenomena. The solar wind, which strikes the lunar surface unhindered due to the absence of global lunar magnetic field, is capable of triggering plasma instability around the lunar exosphere.

## 2. Plasma Instabilities

Plasma instability is a physical phenomenon which occurs in plasma when an energy build-up takes place there taking the plasma system away from the equilibrium. This energy build can lead to the plasma particle loss either by the generation of plasma waves or the particle beams or deformations in the stable plasma structures. It is, however, to be noted that the total energy of the system is conserved as the instabilities only transfer the free energy from one place to another. The free energy in a plasma system can be in the form of beam kinetic energy, pressure gradients in an effective potential (a light fluid supporting a heavier fluid), anisotropies in the velocity distribution (particle kinetic energy), energy stored in a magnetic field, Spatio-temporal variation in the electric field, etc.

Plasma instabilities are often associated with plasma waves supported by the plasma system. The linear phase of a plasma instability is initiated by a set of waves which are unstable and have an exponential growth rate. Out of all the unstable plasma modes, the one having the highest growth rate will drive the plasma dynamics and structure with its periodicity and location in plasma. This dominant mode is defined by the plasma parameters and all other unstable modes will generate fluctuations around the dominant mode.

Plasma instabilities can be classified in many ways which are as follows [Melzani, 2014]:

1. Instabilities with names: Buneman instability, Weibel instability, Rayleigh-Taylor (RT) instability, Kelvin-Helmholtz (K-H) instability, etc.

2. Theory Instabilities: Ideal MHD instability, Resistive plasma instability, Kinetic instability, etc.

3. Instabilities with physical configuration: Two-stream instability, Bump-in-the-tail instability, Filamentation instability, Kink instability, Sausage instability, Ballooning instability, Tearing instability, etc.

4. Instabilities with plasma waves: Ion acoustic instability, Mirror-mode instability, Ion cyclotron instability, Electron-heat-flux instability, etc.

Apart from these there are some other plasma instabilities such as resonant and non-resonant instabilities, micro- and macro instabilities, etc.

## 3. Instability Analysis in Lunar Plasma Environment and Results

Plasma (Two-stream) instability is analysed in the lunar ionosphere where the solar-wind (electron beam) interacts with the lunar background plasma containing electrons. The ion number density being low in both solar wind as well as the lunar ionosphere is neglected here.

The conditions for the instabilities to generate are analyzed from the plasma instability dispersion relation. The plasma conditions may not be always suitable for the instabilities to evolve as the instability occurrence involves the solar wind electron number density, the solar wind velocity and the background plasma electron number density. Favorable values of these three quantities give instabilities at various lunar altitudes. Once the instability is generated, an attempt shall be made for the estimation of the growth rates of these instabilities which in turn will provide information regarding the temporal evolution of these instabilities. The dielectric function of the beam-plasma system for various values of $\alpha$ [Piel, 2010] is shown in Figure 1.

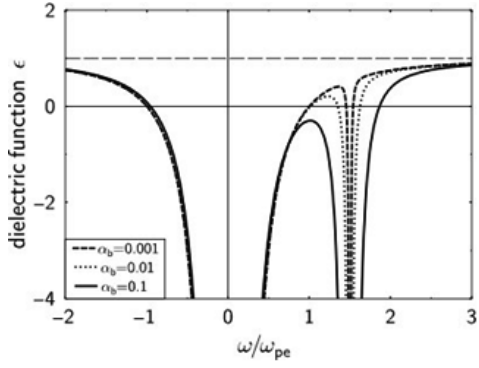

Figure 1: Beam-plasma dielectric function [Piel, 2010]. For a given value of the plasma and the solar wind (beam) parameters, if a $k$ is above the threshold then there will be four real solutions. On decreasing, when the $k$ goes below the threshold, the maxima go below zero and the left beam mode and right plasma mode merged and can be seen in Figure 1.

The dispersion relation obtained from the analysis of beam plasma instability is [Anderson et al., 2001]

$$\frac{\omega_p^2}{\omega^2} + \frac{\omega_b^2}{(\omega - \vec{k}.\vec{v_{b0}})^2} = 1 \quad (1)$$

$$D(\omega, \vec{k}) = 1 - \frac{\omega_p^2}{\omega^2} - \frac{\omega_b^2}{(\omega - \vec{k}.\vec{v_{b0}})^2} \quad (2)$$

The term $\vec{k}.\vec{v_{b0}} = \omega_d$ (Doppler shifted frequency) and

$$\frac{n_b}{n_e} = \frac{\omega_b^2}{\omega_p^2} = \alpha \quad (3)$$

Hence, the dispersion relation becomes

$$D(\omega, \vec{k}) = 1 - \frac{\omega_p^2}{\omega^2} - \frac{\alpha \omega_p^2}{(\omega - \omega_d)^2} \quad (4)$$

The local extremum of this function is obtained

$$\frac{dD}{d\omega} = -\omega_p^2 \frac{-2}{\omega^3} - \alpha \omega_p^2 \frac{-2}{(\omega - \omega_d)^3} = 0 \quad (5)$$

which becomes

$$(\omega - \omega_d)^3 + \alpha \omega^3 = 0 \quad (6)$$

and gives

$$\left(\omega - \omega_d + \alpha^{1/3}\omega\right)\left[(\omega - \omega_d)^2 + \left(\alpha^{1/3}\omega\right)^2 - \left(\alpha^{1/3}\omega\right)(\omega - \omega_d)\right] = 0 \quad (7)$$

Case 1: When 1st term is zero in equation (7)

$$\omega\left(1 + \alpha^{1/3}\right) - \omega_d = 0$$

Now, $\omega = \omega_m$ when $dD/d\omega = 0$

$$\omega_m = \frac{\omega_d}{(1 + \alpha^{1/3})} \quad (8)$$

If we put back the expression of alpha then it becomes

$$\omega_m = \frac{\omega_d \omega_p^{2/3}}{\left(\omega_p^{2/3} + \omega_b^{2/3}\right)} \quad (9)$$

This is the local maxima of the function

Case 2: When 2nd term is zero in above equation

$$\left(1 + \alpha^{2/3} + \alpha^{1/3}\right)\omega^2 + \left(-2\omega_d - \alpha^{1/3}\omega_d\right)\omega + \omega_d^2 = 0$$

$\omega$ is obtained from this quadratic equation as:

$$\omega = \frac{\left(2 + \alpha^{1/3}\right)\omega_d \pm i\sqrt{3}\alpha^{1/3}\omega_d}{2(1 + \alpha^{2/3} + \alpha^{1/3})}$$

Here, the solution is complex unable to give the local extremum of the dispersion function. Hence, the 1st case real solution is considered to find the local maxima as

$$D_m(\omega_m, \vec{k}) = 1 - \frac{\omega_p^2}{\omega_m^2} - \frac{\alpha \omega \omega_p^2}{(\omega_m - \omega_d)^2} \quad (10)$$

Substituting $\omega_m$ from (8), we get

$$D_m(\omega_m, \vec{k}) = 1 - [\frac{\omega_p^2}{\omega_m^2} + \frac{\alpha \omega_p^2}{\omega_d^2}\frac{1}{\alpha^{2/3}}](1 + \alpha^{1/3})^2 \quad (11)$$

This local maxima determines when instability is generated. The maxima depend on the solar wind and lunar plasma parameters. Two cases exist here, one where $D_m > 0$ and the other when $D_m < 0$.

### 3.1 $D_m > 0$: Real roots

When $D_m > 0$, four real roots of the dispersion relation are obtained as follows ($K = e^2/m\varepsilon_0$):

$$D(\omega, \vec{k}) = 1 - \frac{Kn_e}{\omega^2} - \frac{K\alpha n_e}{(\omega - \omega_d)^2} = 0 \quad (12)$$

which gives

$$\omega^4 - (2\omega_d)\omega^3 + (\omega_d^2 - Kn_e - K\alpha n_e)\omega^2 + (2\omega_d Kn_e)\omega - Kn_e\omega_d^2 = 0 \quad (13)$$

It is evident from this 4th order quartic equation in $\omega$ terms that there are two singularities at the 2nd and 3rd term. The D ($\omega$, $k$) expression gives singularity at $\omega = 0$ due to the space charge of lunar plasma. This is the plasma singularity and the modes (normal plasma space charge oscillations) around this singularity are also solar wind independent. The other singularity is at $\omega = \omega_d$ which occurs only when there is a beam (the solar wind) in the system irrespective of its strength (the number density and velocity).

The value of beam singularity $\omega_d = kv_b$ is large as $v_b$ is of the order of $10^5$ m/sec (solar wind velocity) and is located far away from the origin. The nature of dispersion relation remains independent of the singularity $\omega_d = kv_b$ and also around the 2nd singularity (one singularity doesn't affect the nature around another singularity). So, it can be split into two parts - one containing the plasma singularity and the other containing the beam modes and solved separately for the four propagating plasma modes.

$$D(\omega, \vec{k}) = 1 - \frac{Kn_e}{\omega^2} - \frac{K\alpha n_e}{(\omega - \omega_d)^2} = 0$$

Near $\omega \approx 0$, $1 - \omega_p^2/\omega^2 = 0$ giving 2 roots (plasma modes)

$$\omega = \omega_p = \pm\sqrt{\frac{e^2 n_e}{m\epsilon_0}}$$

Near $\omega \approx \omega_d$

$$1 - \frac{K\alpha n_e}{(\omega - \omega_d)^2} = 0$$

which also gives 2 roots which are called beam modes

$$\omega = \omega_d \pm \sqrt{K\alpha n_e}$$

$$\omega = \omega_d \pm \sqrt{\frac{e^2 \alpha n_e}{m\epsilon_0}}$$

The phase velocities of these two beam modes are

$$v_\phi = \frac{\omega}{k} = \frac{\omega_d}{k} \pm \frac{1}{k}\sqrt{\frac{e^2 \alpha n_e}{m\epsilon_0}}$$

$$v_\phi = \frac{\omega}{k} = v_{b0} \pm \frac{1}{k}\sqrt{\frac{e^2 \alpha n_e}{m\epsilon_0}}$$

The two beam modes are travelling waves with different phase velocities but same group velocities which imply these modes are superposition of many waves. Also, the group velocity is equal to the beam velocity which suggests that the beam modes are in resonance with the beam itself and this can lead to an interaction between the modes and the beam.

The other two plasma modes have both zero phase and group velocity, because they are plasma oscillations. These modes can become plasma waves if the thermal velocities of plasma constituents are taken into account in the analysis which is not doing here assuming 'cold' lunar plasma.

In summary of the propagating modes, two beam (solar wind) independent solutions are obtained and these two exist due to the lunar plasma and the other two modes (travelling waves) comes into existence only when the solar wind with high velocities comes and interacts with the lunar plasma.

### 3.2 $D_m < 0$: Complex roots

The instability will be seen only if $D_m \leq 0$ and this condition give a threshold for the plasma instability generation as

$$D_m(\omega_m, \vec{k}) = 1 - \left[\frac{\omega_b^2}{\omega_m^2} + \frac{\alpha\omega_b^2}{\omega_d^2}\frac{1}{\alpha^{2/3}}\right]\left(1 + \alpha^{1/3}\right)^2 \leq 0$$

Now,

$\frac{n_b}{n_e} = \frac{\omega_b^2}{\omega_p^2} = \alpha$, $\omega_p^2 = \frac{n_e e^2}{m\epsilon_0}$, $\omega_b^2 = \frac{n_b e^2}{m\epsilon_0}$, $\omega_d = \vec{k}.\overrightarrow{v_{b0}}$

Substituting these values in the above relation

$$\frac{1}{k^2} \geq \frac{v_{ob}^2 m\epsilon_0}{e^2\left[1 + \left(\frac{n_b}{n_e}\right)^{1/3}\right]^2\left[n_e + n_b\frac{1}{\left(\frac{n_b}{n_e}\right)^{2/3}}\right]}$$

The threshold for lunar instability generation is given as

$$k^2 \leq \frac{e^2\left[1 + \left(\frac{n_b}{n_e}\right)^{1/3}\right]^2\left[n_e + n_b\frac{1}{\left(\frac{n_b}{n_e}\right)^{2/3}}\right]}{v_{ob}^2 m\epsilon_0} \quad (14)$$

It can be inferred here that for the given values of lunar plasma density $n_e$, the solar wind (beam) number density $n_b$ and solar wind (beam) velocity $v_b$, an instability will always get triggered.The lunar plasma number density is typically in range 10 cm$^{-3}$ to 300 cm$^{-3}$and the solar wind parameters are $n_b = 5$ cm$^{-3}$ and $v_b = 3.75 \times 10^5$m/sec.

At any given lunar altitude where the value of these parameters is known, the propagation constants can be obtained for which the instability gets generated. Hence, at a specific altitude, multiple instabilities of different $k$ can be triggered leading to the generation of multiple plasma waves.

## 4. Lunar Plasma Instability: Results
### 4.1 Growth Factor
When the local maxima of $D(\omega, k)$ goes below zero then the two roots vanish, and only a single plasma and beam mode survive. These two roots are complex conjugates of each other ($\omega = \omega_r + \omega_{im}$) and are obtained by Taylor expansion of the dispersion function about the maxima.
Expanding $D(\omega,k)$ about the local maxima $\omega_m$ as given in equation (8)

$$D(\omega) = D(\omega_m) + (\omega - \omega_m)\frac{dD}{d\omega}\Big|\omega = \omega_m +$$
$$\frac{1}{2!}(\omega - \omega_m)^2\frac{d^2D}{d\omega^2}\Big|\omega = \omega_m + \cdots \quad (15)$$

Now, $D(\omega_m)$ is given by equation (11), $D(\omega_m) < 0$ which is the threshold and $D'(\omega_m) = 0$ since this point is local maxima and also $D''(\omega_m) < 0$. The required dispersion relation is $D(\omega) = 0$ which is a quadratic equation in terms of $(\omega - \omega_m)$

$$D(\omega_m) + (\omega - \omega_m)D'(\omega_m) + \frac{1}{2}(\omega - \omega_m)^2 D''(\omega_m) = 0$$

$\omega - \omega_m = \frac{-D'(\omega_m) \pm \sqrt{(D'(\omega_m))^2 - 2D''(\omega_m)D(\omega_m)}}{D''(\omega_m)} = 0$

But $D'(\omega_m) = 0$, hence

$$\omega = \omega_m \pm i\sqrt{\frac{2D(\omega_m)}{D''(\omega_m)}}$$

Both quantities inside the root are positive, so by Taylor expansion, the dispersion relation about the maxima gives two complex conjugate roots for $\omega$.The real part of $\omega$ is given as $\omega_{re}$

$$\omega_{re} = \omega_m = \frac{\omega_d}{(1 + \alpha^{1/3})} = \frac{\omega_d\omega_p^{2/3}}{\left(\omega_p^{2/3} + \omega_b^{2/3}\right)} \quad (16)$$

And it gives the phase and group velocities of instability

$$v_p = \frac{\omega_{re}}{k} = \frac{v_b}{\left(1 + \alpha^{1/3}\right)}$$

$$v_g = \frac{d\omega_{re}}{dk} = \frac{v_b}{\left(1 + \alpha^{1/3}\right)}$$

In these expressions, when $\alpha \ll 1$, there is resonance between the instability's propagation and the movement of the solar wind (beam) making energy transfer from electrons to the instability.

The imaginary part of $\omega$ is the growth rate (amplitude increase with time)

$$\omega_{im} = \sqrt{\frac{2D(\omega_m)}{D''(\omega_m)}} \quad (17)$$

Now, the term $D''(\omega_m)$ is obtained from equation (5) as

$$\frac{d^2D}{d\omega^2} = -6\left[\frac{\omega_p^2}{\omega^2} + \frac{\omega_b^2}{(\omega - \omega_e)^4}\right] \quad (18)$$

At $\omega = \omega_m$, and using equation (8), equation (18) becomes

$$\frac{d^2D}{d\omega^2} = -6\frac{\left(1 + \alpha^{1/3}\right)^4}{\omega_d^4}\left[\omega_p^2 + \frac{\omega_b^2}{\alpha^{4/3}}\right] \quad (19)$$

$D(\omega_m)$ and $d^2D/d\omega^2$ are inserted in equation (17) to obtain $\omega_{im}$

$$= \frac{\omega_d}{\sqrt{3}\left(1 + \alpha^{1/3}\right)^2}\alpha^{1/3}\left[\frac{\omega_b^2\left(1 + \alpha^{1/3}\right)^2 + \omega_p^2\alpha^{2/3}\left(1 - \alpha^{1/3}\right)^2 - \omega_d^2\alpha^{2/3}}{\omega_b^2 + \omega_p^2\alpha^{4/3}}\right]^{1/2}$$

Dividing the whole by $\omega_p^2$ and taking $\omega_b^2/\omega_p^2 = \alpha$, it becomes

$$\omega_{im} = \frac{\alpha^{1/6}\omega_p}{\sqrt{3}\left(1 + \alpha^{1/3}\right)^{5/2}}\left(\frac{\omega_d^2}{\omega_p^2}\right)\left[\frac{\omega_p^2}{\omega_d^2}\left(1 + \alpha + 3\alpha^{1/3} + 3\alpha^{2/3}\right) - 1\right]^{1/2} \text{- (20)}$$

This expression shows that that the growth factor depends on lunar plasma, solar wind (beam) parameters and the $k$ of plasma wave. At a given lunar altitude, multiple different instability modes can exist which are differentiated by their $k$ as per the threshold. Every such mode with a specific $k$ will have its growth rate given by equation (20).

The growth factor with $k$ (for different values of $n_e$) is plotted in Figure 2 to find the fastest growing mode and to look for the expression for a specific $k$ which gives the maximum growth. The $n_e$ variation is taken as 10 cm$^{-3}$, 50 cm$^{-3}$, 100 cm$^{-3}$, 200 cm$^{-3}$ and 300 cm$^{-3}$.

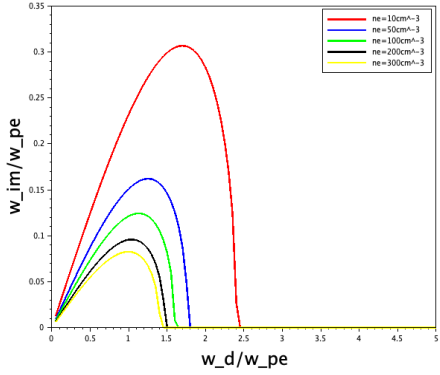

Figure 2: The growth factor with k for different $n_e$.

Let $K = (1 + \alpha + 3\alpha^{1/3} + 3\alpha^{2/3})$, equation (20) becomes

$$\omega_{im} = \frac{\alpha^{1/6}\omega_p}{\sqrt{3}\left(1+\alpha^{1/3}\right)^{5/2}}\left(\frac{\omega_d^2}{\omega_p^2}\right)\left[\frac{\omega_p^2}{\omega_d^2}K - 1\right]^{1/2} \text{-(21)}$$

$$\frac{d\omega_{im}}{d\omega_d} = \frac{\alpha^{1/6}\omega_p}{\sqrt{3}\left(1+\alpha^{1/3}\right)}\left[\frac{2\omega_d}{\omega_p^2}\left(\frac{\omega_p^2 K}{\omega_d^2}-1\right)^{1/2}\right.$$
$$\left. - \frac{\omega_d^2}{\omega_p^2}\frac{\omega_p^2 K}{\omega_d^2}\left(\frac{\omega_p^2 K}{\omega_d^2}-1\right)^{-1/2}\right]$$

For maxima $d\omega_{im}/d\omega_d = 0$

$$\frac{\alpha^{1/6}\omega_p}{\sqrt{3}\left(1+\alpha^{1/3}\right)}\left[\frac{2\omega_d}{\omega_p^2}\left(\frac{\omega_p^2 K}{\omega_d^2}-1\right)^{1/2}\right.$$
$$\left. - \frac{\omega_d^2}{\omega_p^2}\frac{\omega_p^2 K}{\omega_d^2}\left(\frac{\omega_p^2 K}{\omega_d^2}-1\right)^{-1/2}\right] = 0$$

which gives

$$\frac{\omega_p^2 K}{\omega_d^2} - 1 = 0, \omega_d = \pm\omega_p\sqrt{K}$$

Now $\omega_d = kv_b$, hence we get

$$k = \frac{\omega_d}{v_b} = \frac{\omega_p\sqrt{K}}{v_b}$$

Inserting $K$ from above

$$k_{max} = \left(\frac{\omega_p}{v_b}\right)\sqrt{1 + \alpha + 3\alpha^{1/3} + 3\alpha^{2/3}}$$

### 4.2 Decay Rate

When the dispersion relation is expanded in the Taylor series, the complex solutions comes

$$\omega = \omega_m \pm i\sqrt{\frac{2D(\omega_m)}{D''(\omega_m)}}$$

This gives two solutions which are complex conjugates of each other. The imaginary partgives the beam plasma instability and the other a propagating mode.
$E = E_0 e^{i(kx-\omega t)}$ , $\omega = \omega_r \pm i\omega_{im}$, $E = E_0 e^{i(kx-[\omega_r \pm i\omega_{im}]t)}$
which becomes

$$E = E_0 e^{\pm\omega_{im}t}e^{i(kx-\omega_r t)}$$

The solution corresponding to $+\omega_{im}$ correspond to the instability and the solution with $-\omega_{im}$ corresponds to a decaying mode as with time the amplitude will decrease.

These growing and decaying modes show striking similarity in its properties with the instabilities. The decay rate with respect to $k$ value of the modes for different values of $n_e$ is shown in Figure 3. It can be seen from the figure that a mode with highest decay rate will decay early and with increasing $n_e$, the $k$ value with the highest decay rate will also increase.

Therefore, it can be concluded that the solar wind impact on the lunar plasma can trigger (i) Langmuir plasma oscillations; (ii) Right beam propagating modes; (iii) Beam-plasma instability; (iv) Beam plasma decay mode.

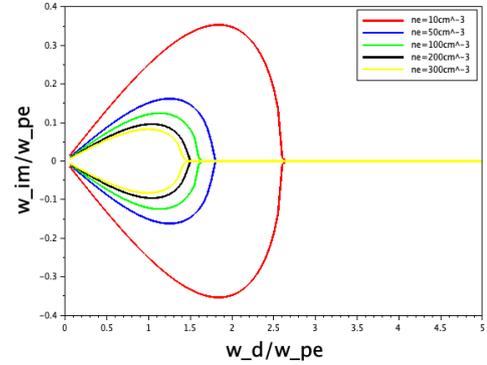

Figure 3: Symmetric nature of decay and growth rates.

The growth rate and decay rate are symmetric to each other which means for a given $n_e$, the maximum growth rate and decay rate occur at the same $k$ value. This strong symmetric nature suggests a strong energy exchange mechanism i.e. with time the instability modes gain energy and in other case with time the decay modes loose energy. It suggests that a resonance exist between the solar wind, lunar plasma and plasma instability where during the instability growth the solar wind dump energy in the lunar plasma and during the decay the lunar plasma pumps energy in the solar wind quenching the instability.